\documentclass[a4paper]{article}

\usepackage{INTERSPEECH2021}

\usepackage[utf8]{inputenc}

\usepackage{tipa}

\usepackage{url}

\title{Adaptation of Tacotron2-based Text-To-Speech for \\ Articulatory-to-Acoustic Mapping using Ultrasound Tongue Imaging}
\name{Csaba Zainkó$^{1}$, László Tóth$^{2}$, Amin Honarmandi Shandiz$^{2}$, Gábor Gosztolya$^{3}$, Alexandra Markó$^{4,5}$, Géza Németh$^{1}$, Tamás Gábor Csapó$^{1,5}$}
\address{
  $^1$Department of Telecommunications and Media Informatics, \\
	Budapest University of Technology and Economics, Budapest, Hungary \\
	$^2$Institute of Informatics, University of Szeged, Hungary\\
	$^3$MTA-SZTE Research Group on Artificial Intelligence, Szeged, Hungary \\
	$^4$Department of Applied Linguistics and Phonetics, Eötvös Loránd University, Budapest, Hungary \\
	$^5$MTA-ELTE Lendület Lingual Articulation Research Group, Budapest, Hungary}
\email{\{zainko, nemeth, csapot\}@tmit.bme.hu, \{tothl, shandiz, ggabor\}@inf.u-szeged.hu, marko.alexandra@btk.elte.hu}

\begin{document}

\maketitle
\begin{abstract}
For articulatory-to-acoustic mapping, typically only limited parallel training data is available, making it impossible to apply fully end-to-end solutions like Tacotron2. In this paper, we experimented with transfer learning and adaptation of a Tacotron2 text-to-speech model to improve the final synthesis quality of ultrasound-based articulatory-to-acoustic mapping with a limited database. We use a multi-speaker pre-trained Tacotron2 TTS model and a pre-trained WaveGlow neural vocoder. The articulatory-to-acoustic conversion contains three steps: 1) from a sequence of ultrasound tongue image recordings, a 3D convolutional neural network predicts the inputs of the pre-trained Tacotron2 model, 2) the Tacotron2 model converts this intermediate representation to an 80-dimensional mel-spectrogram, and 3) the WaveGlow model is applied for final inference. This generated speech contains the timing of the original articulatory data from the ultrasound recording, but the F0 contour and the spectral information is predicted by the Tacotron2 model. The F0 values are independent of the original ultrasound images, but represent the target speaker, as they are inferred from the pre-trained Tacotron2 model. In our experiments, we demonstrated that the synthesized speech quality is more natural with the proposed solutions than with our earlier model.
	
\end{abstract}

\section{Introduction}
\label{sec:intro}

Articulatory-to-acoustic mapping (AAM) methods aim to synthesize the speech signal directly from articulatory input, as opposed to text-to-speech, when speech is synthesized from the textual input. AAM applies the theory that articulatory movements are directly linked with the acoustic speech signal in the speech production process. A recent potential application of this mapping is a “Silent Speech Interface” (SSI~\cite{Denby2010,Schultz2017a,Gonzalez-Lopez2020}), which has the main idea of recording the soundless articulatory movement, and automatically generating speech from the movement information, while the subject does not produce any sound. Such an SSI system can be highly useful for the speaking impaired (e.g. after laryngectomy or elderly people), and for scenarios where regular speech is not feasible, but the information should be transmitted from the speaker (e.g. extremely noisy environments or military applications).

For the articulatory-to-acoustic mapping, the typical input can be electromagnetic articulography (EMA)~\cite{Cao2018,Taguchi2018}, ultrasound tongue imaging (UTI)~\cite{Denby2004,Hueber2010,Hueber2011,Jaumard-Hakoun2016,Tatulli2017,Csapo2017c,Grosz2018,Toth2018,Moliner2019,Gosztolya2019,Csapo2019,Csapo2020c,Toth2020,Shandiz2021}, permanent magnetic articulography (PMA)~\cite{Gonzalez2017a,Gonzalez-Lopez2021}, surface electromyography (sEMG)~\cite{Janke2017,Diener2018a}, Non-Audible Murmur (NAM)~\cite{Shah2018}, electro-optical stomatography~\cite{Stone2016}, impulse radio ultra-wide band (IR-UWB)~\cite{Shin2016}, radar~\cite{Digehsara2021} or video of the lip movements~\cite{Hueber2010,Ephrat2017,Sun2018}.
From another aspect, there are two distinct ways of SSI solutions, namely `direct synthesis' and `recognition-and-synthesis'~\cite{Schultz2017a}. In
the first case, the speech signal is generated without an intermediate step, directly from the articulatory data~\cite{Cao2018,Taguchi2018,Denby2004,Hueber2011,Jaumard-Hakoun2016,Csapo2017c,Grosz2018,Moliner2019,Gosztolya2019,Csapo2019,Gonzalez2017a,Janke2017,Diener2018a,Shah2018,Ephrat2017}. In the second case, silent speech recognition (SSR) is applied on the biosignal which extracts the content spoken by the person (i.e. the result of this step is text); this step is then followed by text-to-speech (TTS) synthesis~\cite{Hueber2010,Tatulli2017,Toth2018,Stone2016,Sun2018,Arthur2021}. 
In the SSR+TTS approach, any information related to speech prosody is lost, whereas it may be kept with direct synthesis. Also, the smaller delay by the direct synthesis approach might enable conversational use.

For the direct conversion, typically, vocoders are used, which synthesize speech from the spectral parameters predicted by the DNNs from the articulatory input. One of the spectral representations that was found to be useful earlier for statistical parametric speech synthesis is Mel-Generalized Cepstrum in Line Spectral Pair form (MGC-LSP)~\cite{Csapo2015d,Csapo2016}. 
Since the introduction of WaveNet in 2016~\cite{Oord2016}, neural vocoders can generate highly natural raw samples of speech, conditioned on mel-spectrogram or other input. One of the most recent types of neural vocoders, WaveGlow~\cite{Prenger2019} is a flow-based network capable of generating high-quality speech from mel-spectrograms. The advantage of the WaveGlow model is that it is relatively simple, yet the synthesis can be done faster than real-time. In~\cite{Csapo2020c}, we integrated the WaveGlow neural vocoder into ultrasound-based articulatory-to-acoustic conversion.

In the latest years, most TTS solutions apply end-to-end methods, by operating directly on character or phoneme input sequences and producing raw speech signal outputs. One of the most widely used solutions is Tacotron2~\cite{Shen2018}, which applies a recurrent sequence-to-sequence feature prediction network that maps character embeddings to mel-scale spectrograms, followed by a neural vocoder. The encoder-decoder network, using the attention mechanism, 
encodes a specific attribute of speech and maps sequences of differing length.
In~\cite{Shen2018}, the input characters are represented with a learned 512-dimensional embedding, which ensures that traditional text processing is not necessary on the input.


In the field of AAM, according to our knowledge, only a few studies have used fully end-to-end / sequence-to-sequence solutions~\cite{Zhang2021b,Mira2021}. Zhang and his colleagues introduced TaLNet, which is based on an encoder-decoder architecture, using the attention mechanism. Both ultrasound and lip are used as the input of AAM, from English speakers of the UltraSuite-TaL database~\cite{Ribeiro2021}. First, a Tacotron2 model is trained with a large amount of speech data, and after that, transfer learning is applied with the articulatory input. The presented approach was found to be significantly better than earlier baselines. In the study, they also checked the contribution of each articulatory input, and found that the weakest results could be achieved with the lip-only system, followed by ultrasound-only. The combination of ultrasound and lip (TaLNet) was found to be the best, suggesting that these two modalities complement each other well. In another study, by Mira and his colleagues, end-to-end video-to-speech synthesis was proposed, using GANs~\cite{Mira2021}. The video of the face is translated directly to speech, without an intermediate representation, applying an encoder-decoder architecture. They experimented on various databases and show that the choice of adversarial loss is a key for realistic results.

In this paper, we experiment with transfer learning and adaptation of a Tacotron2 text-to-speech model to improve the final synthesis quality of ultrasound-based articulatory-to-acoustic mapping with a limited database.



\section{Methods}

\begin{figure}
\centering
\includegraphics[trim=0.5cm 3.5cm 0.5cm 3.0cm, clip=true, width=\columnwidth]{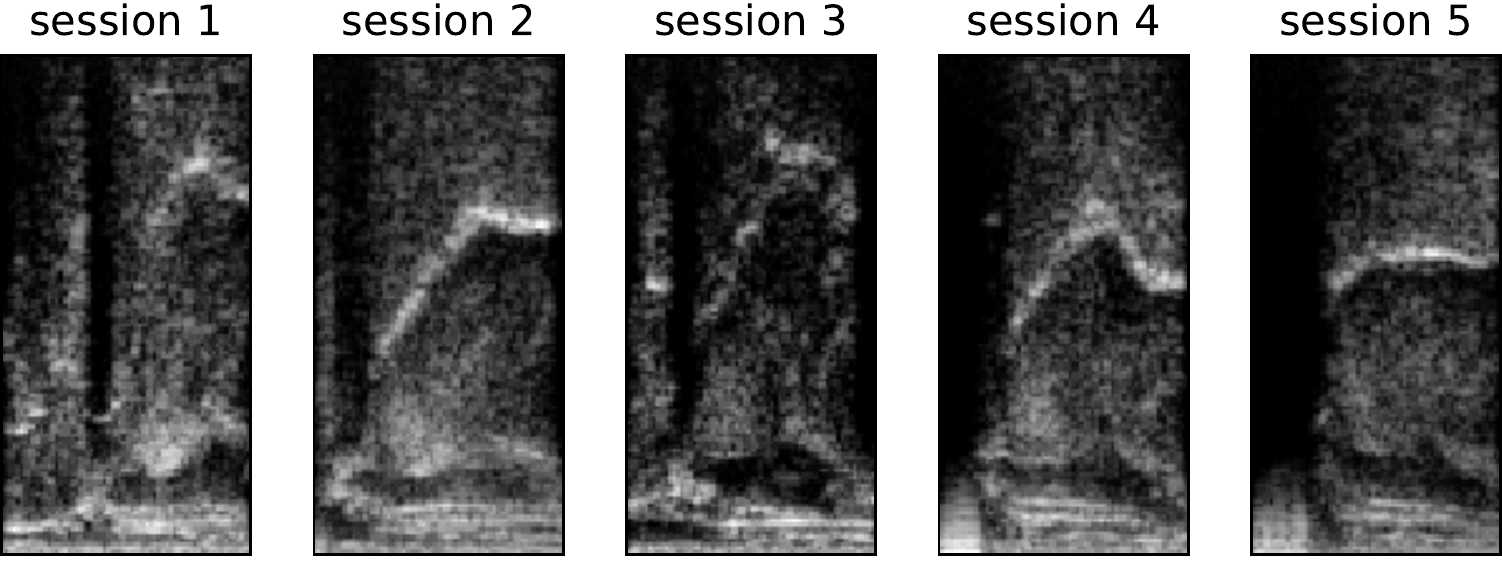}
\vspace{-2mm}
\caption{Sample ultrasound images from the five sessions.}
\label{fig:UTI_sample}
\vspace{-2mm}
\end{figure}

\subsection{Data}

For Tacotron2 and WaveGlow training, we chose 5 male and 6 female Hungarian speakers (altogether 23k sentences, roughly 22 hours) from the PPSD database~\cite{Olaszy2013}. This data served as the acoustic-only training material required for the encoder-decoder architecture and the neural vocoder.

For the articulatory data, we used the Hungarian parallel ultrasound and speech dataset that we recorded for earlier studies~\cite{Csapo2019,Csapo2020c,Gosztolya2020}. We selected a female speaker (speaker048), who was recorded in five sessions (once 209 sentences, and four times 59 sentences). The tongue movement was recorded in midsagittal orientation using the ``Micro'' ultrasound system of Articulate Instruments Ltd. at 81.67 fps. The speech signal was recorded with a Beyerdynamic TG H56c tan omnidirectional condenser microphone. The ultrasound data and the audio signals were synchronized using the tools provided by Articulate Instruments Ltd. In our experiments, the raw scanline data of the ultrasound was used as input of the networks, after being resized to 64$\times$128 pixels using bicubic interpolation (see samples in Fig.~\ref{fig:UTI_sample}), as we found earlier that this reduction does not cause significant information loss~\cite{Csapo2021}. 

For the Tacotron2 speaker adaptation, speaker048's data was used (train: 318 sentences, and validation: 40 sentences).


\subsection{Ultrasound-to-Melspectrogram using 3D-CNN \\ (baseline)}

When we are dealing with image processing as input data, then convolutional neural networks are one of the most popular and effective methods which can extract complex features from data by adding deep layers~\cite{Krizhevsky2012}. In Silent Speech Interface, when we have ultrasound data as input, our input is not only just images but sequences of images which could be considered as a video. Standard CNN considers 2D images to extract features by convolving 2D filters over images. Therefore, to model temporal information, a third dimension has to be considered~\cite{Ji2012,Hochreiter1997}. Recurrent Neural Networks such as Long Short Term Memory (LSTM) are good examples of combining features extracted from both temporal and spatial parts of data~\cite{Hochreiter1997}. Using LSTM networks have some drawbacks such as training difficulties, while some variants of these networks were proposed to mitigate this problem, such as quasi-recurrent neural networks~\cite{Bradbury2016}. 

Here we use another variation by adding a third dimension as (2+1)D CNN which shows good performance in video action recognition task~\cite{Tran2018}. It shows good results when used with ultrasound images and it could be considered as a substitute of CNN+LSTM~\cite{Toth2020}. In the baseline system of the current study, we apply the same 3D CNN which was used in~\cite{Toth2020} for predicting 80-dimensional melspectrogram features from ultrasound tongue image input. 

\begin{figure}
\centering
\includegraphics[trim=0.3cm 0.7cm 0.3cm 0.3cm, clip=true, width=\columnwidth]{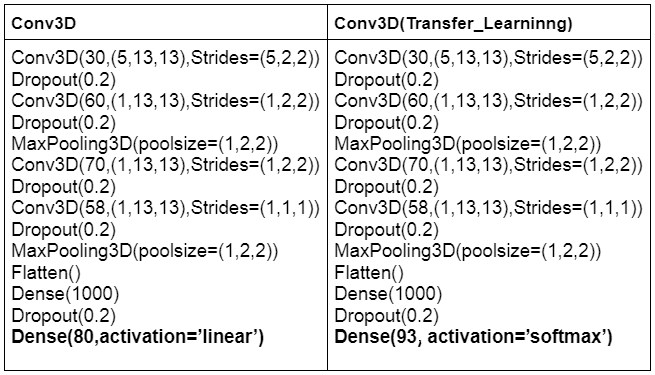}
\vspace{-2mm}
\caption{The layers of the 3D CNNs in the Keras implementation, along with their most important parameters. Left: baseline 3D CNN for melspectrogram prediction, right: proposed 3D CNN for symbol prediction.}
\label{fig:3D-CNN}
\vspace{-2mm}
\end{figure}

This network processed 5 frames of video that were 6 frames apart (6 is the stride parameter of the convolution along the time axis)~\cite{Toth2020}. Following the concept of (2+1)D convolution, the five frames were first processed only spatially, and then got combined along the time axis just below the uppermost dense layer. Fig.~\ref{fig:3D-CNN} left shows the actual network configuration. The training was performed using the SGD optimizer with 0.06 starting learning rate. It was reduced when a validation MSE has stopped improving by factor 0.5. The batch size was 128. The training objective function was the mean squared error (MSE).



\subsection{Ultrasound-to-Symbol using 3D-CNN}
\label{sec:ult-to-sym}

In the proposed system, we use the same structure of the 3D CNN as in the baseline system. The difference is in the target of the network: we predict symbols of Tacotron2 internal representation, having 93 dimensions. At first, we trained with the same methods as the baseline model, but the model was not applicable. We fine-tuned the optimizer, batch size, and other hyperparameters but the model still did not train. Sometimes the accuracy was zero or it learned only the silent symbol and predicted it everywhere. Finally, transfer learning was successful. We reused the baseline 3D-CNN model’s weights at the convolutional layers. All convolutional layers were frozen and only the last two FC layers (with 1000 and 93 neurons) were trained. The weights of these two layers were initialized randomly. Here, cross-entropy is used as the loss function. Because the classes of symbols were not balanced, we used a specific loss function: the loss was weighted with the occurrence of the symbols. We used Adam optimizer and accuracy as a metric. The other parameters of the CNN are the same as the baseline, see Fig.~\ref{fig:3D-CNN} right.

\subsubsection{Accuracy and the confusion matrix}
\label{sec:conf-matrix}

\begin{figure}
\centering
\includegraphics[trim=0.0cm 0.0cm 0.0cm 0.0cm, clip=true, width=1.1\columnwidth]{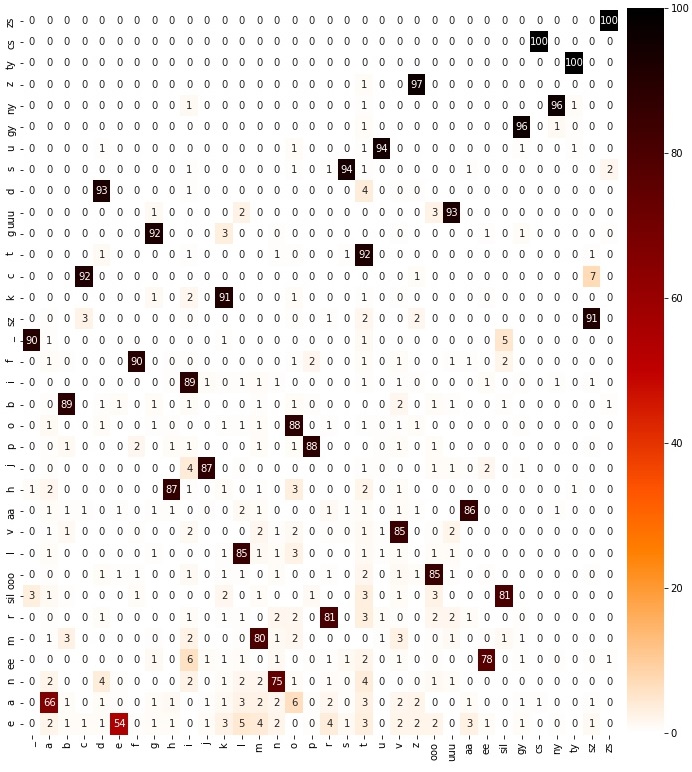}
\vspace{-2mm}
\caption{Simplified confusion matrix of the proposed Ultrasound-to-Symbol 3D-CNN. The values are normalized and showed in percentages. Rows: target, columns: predicted.}
\label{fig:conf}
\vspace{-2mm}
\end{figure}

The Ultrasound-to-Symbol 3D-CNN model reached 0.68 validation accuracy after 20 epochs (train acc.: 0.83). Early stopping was used with a patience parameter of 7. To improve our Tacotron2 model, the confusion matrix was used to generate augmented training data (see later in Sec.~\ref{sec:prop_2}). Fig.~\ref{fig:conf} shows a simplified version of the confusion matrix (for visualization purposes only -- the full matrix involves all 93 symbols: for this figure, we removed the symbols which were not used in the current models and pooled together the short and long versions of the symbols). The values are normalized by rows (target symbols) and converted to percentage values. The first row (on the top) is the most accurate symbol, and the last row (on the bottom) is the least accurate symbol. We expected that the errors are related to  articulation, but in Fig.~\ref{fig:conf} it seems mainly noise-like. The symbols with lower accuracies were some vowels and nasals (e,a,ee,n,m in the figure, /\textipa{E,O,e:,n,m}/ in IPA). The symbols with higher accuracies were some less frequent consonants (z,ty,cs,zs in the figure, /\textipa{Z,tS,c,z}/ in IPA). 

\subsection{Symbol-to-melspectrogram using Tacotron2}


We used a multi-speaker Tacotron2 model~\cite{Shen2018} based on the NVIDIA implementation (\url{https://github.com/NVIDIA/tacotron2}). The speakers’ IDs are coded as a one-hot vector and added to the inputs of the LSTM cells both in the encoder and decoder. The model was trained by all 11 speakers of the PPSD database~\cite{Olaszy2013} at the same time. The order of all speakers’ sentences was randomized. The input of the Tacotron2 is a sequence of symbols. Because Hungarian is an almost phonetic language, we used a mixed collection of letters and phonemes. The symbols of the input sequence follow the phonemes of the sentences, but we did not use allophones or other detailed discrimination. Only the long--short property is used to encode durational differences. The phonemes are represented with their approximate letter: the lowercase letters show the short phonemes, the capital letters indicate the long phonemes. 
 
This multi-speaker model was trained during 156k iterations on a single NVIDIA Titan Xp. The sample rate of the sound was 22\,050~Hz, the window size was 1024 and the hop length was 256. We used 80 mel channels between 0~Hz and 8000~Hz to keep compatibility with the WaveGlow model. The encoder’s symbols embedding and embedding dimension was also 512. The decoder’s RNN dimensions were 1024.

\begin{figure}[t]
\centering
\includegraphics[trim=0.0cm 0.0cm 0.0cm 0.3cm, clip=true, width=0.8\columnwidth]{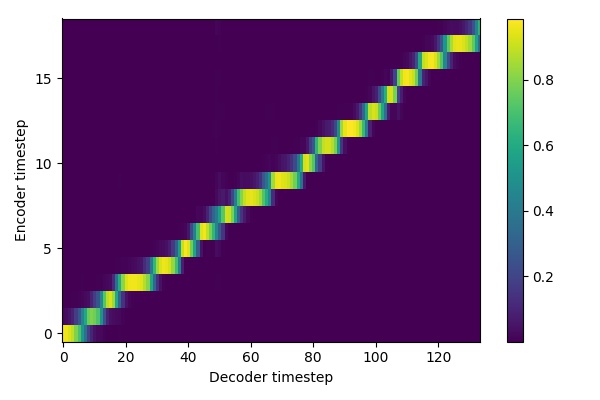}
\includegraphics[trim=0.0cm 0.0cm 0.0cm 0.3cm, clip=true, width=0.8\columnwidth]{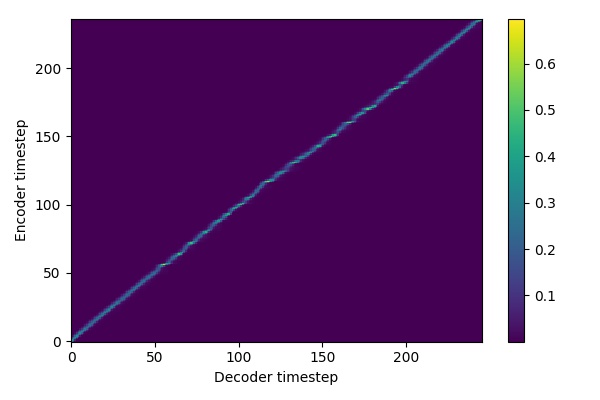}
\includegraphics[trim=0.0cm 0.0cm 0.0cm 0.3cm, clip=true, width=0.8\columnwidth]{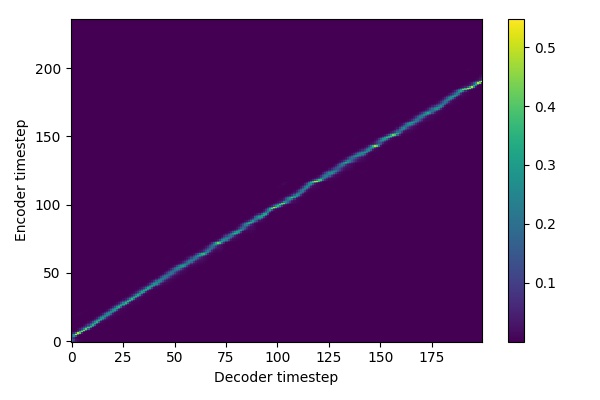}
\vspace{-2mm}
\caption{Examples for the connection between the steps of the encoder and decoder. Top: Tacotron2 without timing information. Middle: Tacotron2 with timing information (Proposed \#1). Bottom: Tacotron2 with timing information and with data augmentation (Proposed \#2).}
\label{fig:alignments}
\vspace{-2mm}
\end{figure}


Our goal was to use our pre-trained Tacotron2 model (originally developed for TTS) without modification, therefore we made only some fine-tuning for AAM purposes. The ultrasound image sequence does not contain F0-related information, but it contains the timing of speech. Basically, the Tacotron2 does not handle timing information of a sentence, it can generate that via an attention mechanism. 
Fig.~\ref{fig:alignments} top shows an example for the connection between the steps of the encoder and decoder with this initial Tacotron2 system. This sentence encoder contains 16 symbols plus two padding symbols at the borders of the sentence. The model generated 134 decoder frames. In this model, one frame is about 11.6ms, so this sentence was about 1.6s long. Clearly, the timings are not modeled well here.

\subsubsection{Time-synchronous Tacotron2 system}
\label{sec:ts_tac}
In order to use the proper timing of the input sequence, we generated a new training set from the original 11 speakers' dataset. The input symbols were repeated accordingly to the real duration of a phone. The repeating number was calculated from the ultrasound frame rate (81.67 fps). For example, at a 98ms long phone, the symbol was repeated 8 times. The attention mechanism adapted to the synchronized input during the fine-tuning. It required 7.5k iterations.

\subsubsection{Proposed system \#1}
\label{sec:prop_1}

The speaker in the ultrasound dataset (speaker048) is independent of the 11 speakers of the training set of Tacotron2. The next step was fine-tuning to the new speaker. We chose a female speaker from the 11 others, and at the tuning, her speakerID one-hot vector was used. At this step, 84 iterations resulted in the smallest validation error. In the first proposed system, this model was used. Fig.~\ref{fig:alignments} middle shows the proper timing of the generated speech. The input of that sentence contains 237 symbols, and the system generated 246 output frames. The difference comes from the uncertainty of the end decision of the decoder. The figure also shows the Tacotron2 can tolerate some symbol errors, i.e.~the line is not perfectly straight; there are some small steps, where the decoder ignores some input symbols.

\subsubsection{Proposed system \#2}
\label{sec:prop_2}

Our experience was that Tacotron2 can tolerate some mistakes in the prediction of the 3D-CNN model (Sec.~\ref{sec:ult-to-sym}), but these mistakes cause audible distortion during the final synthesis. The distribution of the wrong predictions can be characterized by the confusion matrix (Sec.~\ref{sec:conf-matrix}) of the 3D-CNN network. It is not accurate because it does not contain the position information of the mistakes, but it is suitable to generate similar training data for fine-tuning the Tacotron2 model. With the distribution of the symbol’s error, we modified the 11 speakers training set. The symbol changing was based on the distribution but it was randomized. For every sentence, 20 different versions were generated. The output mel-spectrograms were not changed. 4.3k iterations provided the lowest validation error. Fig.~\ref{fig:alignments} bottom shows the tuned model’s connection between the encoder and decoder. There are two differences compared to the middle subfigure. The number of the encoder steps remained the same, but there are fewer decoder steps. The decoder learned to ignore the different types of silence symbols (pad, sil, start\_sil, end\_sil) which were mixed in the predicted symbol sequence. The other difference is that the line is smoother. It shows that a decoder step connects more encoder steps and the model can combine the information of good and bad symbols. 

After that we also repeated the tuning to the speaker from the ultrasound dataset. Here we also generate modified training data with the phoneme errors. The procedure was the same as at the multi-speaker case. At this second step, 182 iterations were required. We used this model in the second proposed system.




\subsection{Melspectrogram-to-speech with a neural vocoder}

Similarly to the original WaveGlow paper~\cite{Prenger2019}, 80 bins were used for mel-spectrogram using librosa mel-filter defaults (i.e.\ each bin is normalized by the filter length and the scale is the same as in HTK, Hidden Markov Model Toolkit). FFT size and window size were both 1024 samples. For hop size, we use the base 256 samples. This 80-dimensional mel-spectrogram served as the training target of the Tacotron2 network.
A WaveGlow model was trained with the Hungarian data (WaveGlow-HU). This latter training was done on a server with eight V100 GPUs, altogether for 635k iterations.
In the synthesis phase, an interpolation in time was not necessary, different from \cite{Csapo2020c}. The ultrasound frame rate was 270 samples, but the differences were compensated by the Tacotron2 model, the output frame rate of the model was 256 samples which is the same as the WaveGlow's hop size.
Finally, the synthesized speech is the result of the inference with the trained WaveGlow-HU model conditioned on the mel-spectrogram input~\cite{Prenger2019}.

\section{Experiments and Results}

After training the above models, we synthesized sentences from the test part of the ultrasound dataset. These sentences have not been used during the training process, neither in the Ultrasound-to-Symbol model, nor in the Tacotron2 training and tuning process. The domain of the texts is also independent of the training and validation dataset: it contains the Hungarian version of 'The North Wind and the Sun'.



\subsection{Subjective listening test}

In order to determine which proposed version is closer to natural speech, we conducted an online MUSHRA-like test~\cite{mushra}. Our aim was to compare the natural sentences with the synthesized sentences of the baseline, the proposed approaches and a lower anchor system (the latter having constant F0 and 2D CNN predicted MGC-LSP, from~\cite{Csapo2020c}). In the test, the listeners had to rate the naturalness of each stimulus in a randomized order relative to the reference (which was the natural sentence), from 0 (very unnatural) to 100 (very natural). We chose nine sentences from the test set of the target speaker. The variants appeared in randomized order (different for each listener). The samples can be found at \url{http://smartlab.tmit.bme.hu/ssw11_tacotron2}.

\begin{figure}
\centering
\includegraphics[trim=0.25cm 0.3cm 0.2cm 0.25cm, clip=true, width=\columnwidth]{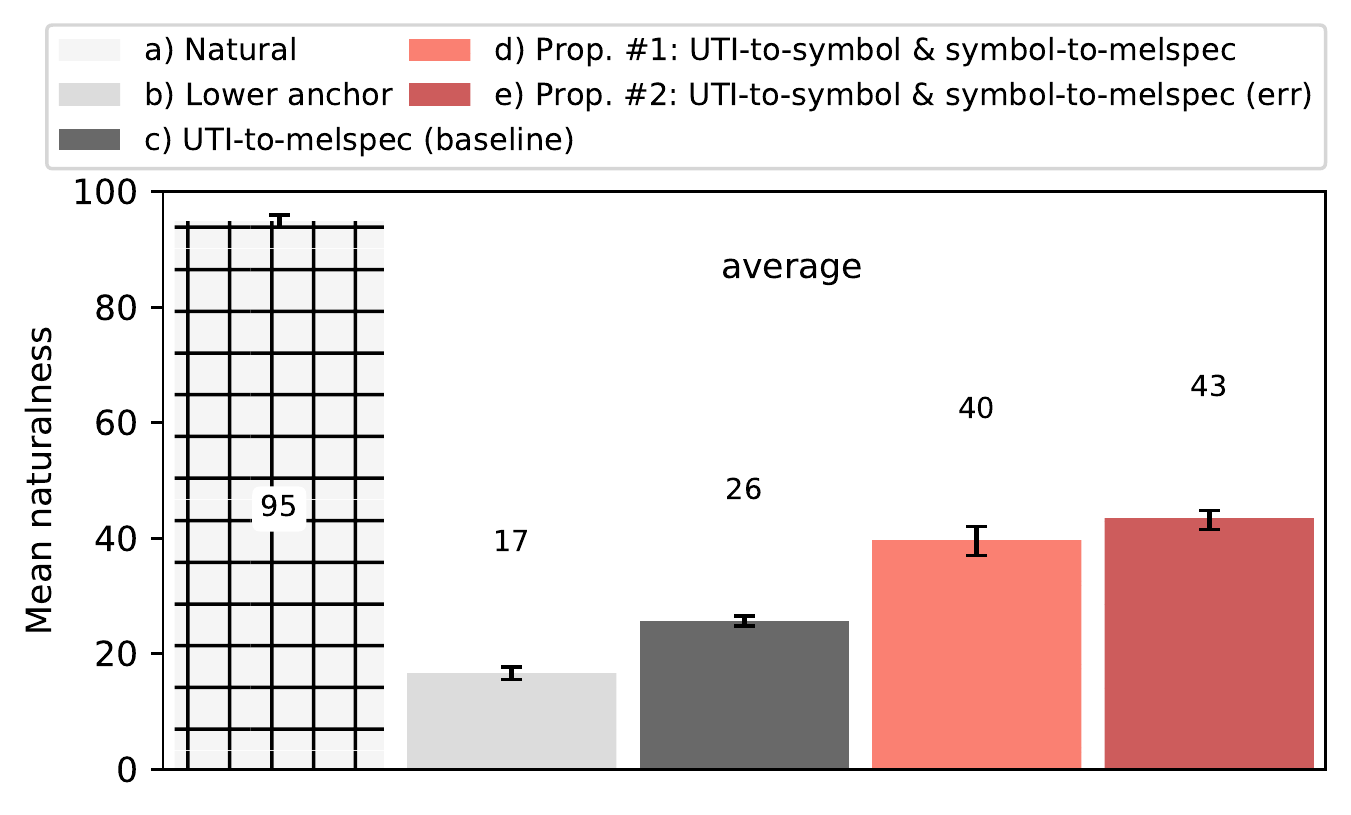}
\vspace{-2mm}
\caption{Results of the subjective evaluation with respect to naturalness. The error bars show the 95\% confidence intervals.}
\label{fig:results_subjective}
\vspace{-2mm}
\end{figure}

Each sentence was rated by 23 native Hungarian speakers (11~females, 12 males; 14--47 years old), in a silent environment. On average, the test took 10 minutes to complete. Fig.~\ref{fig:results_subjective} shows the average naturalness scores for the tested approaches. The lower anchor received the weakest scores, followed by the baseline, and the proposed approaches.
To check the statistical significances, we conducted Mann-Whitney-Wilcoxon ranksum tests with a 95\% confidence level. Based on this, both proposed variants were evaluated as significantly more natural than the baseline. The listeners noted the difference between the two proposed versions: proposed\#1, the one with standard training (Sec.~\ref{sec:prop_1}) was rated as 40\%, while proposed \#2, the one with additional error training (Sec.~\ref{sec:prop_2}) was rated as 43\% -- but this difference is not statistically significant.

As a summary of the listening test, we can conclude that splitting the ultrasound-to-speech prediction task into three parts increased the naturalness, mostly because of the Tacotron2 component which could be trained with a large amount of speech data, and transfer learning / adaptation was possible to the target speaker.

\section{Discussion}

In Sec.~\ref{sec:intro}, we noted that currently only a few sequence-to-sequence / fully end-to-end solutions are available for articulatory-to-acoustic mapping~\cite{Zhang2021b,Mira2021}. Our proposed solution has the following similarities and differences. Mira and his colleagues use the video of the face as input~\cite{Mira2021}, Zhang and his colleagues use both ultrasound and lip video input~\cite{Zhang2021b}, whereas in our study we use ultrasound tongue image input. As the three studies apply different databases, the results are not directly comparable. In~\cite{Mira2021}, GANs are used with specific adversarial loss, whereas we apply 3D CNN to model the spatial and temporal dependencies of the articulatory and acoustic data.
Similarly to~\cite{Zhang2021b}, we apply Tacotron2 as the encoder-decoder network, but we extend the basic training with additional data augmentation, which includes the wrong predictions from the confusion matrix of the UTI-to-symbol prediction network. By using the symbols as intermediate representation, our solution is closer to the 'recognition-and-synthesis' type of SSIs. 



\section{Conclusions}

In this paper, we experimented with transfer learning and adaptation of a Tacotron2 text-to-speech model to improve the final synthesis quality of ultrasound-based articulatory-to-acoustic mapping with a limited database (roughly 200 sentences). We used a Hungarian multi-speaker pre-trained Tacotron2 TTS model and a pre-trained WaveGlow neural vocoder (both trained on 11 speakers's data, altogether 23k sentences, roughly 22 hours of speech). The proposed articulatory-to-acoustic conversion framework is a fully end-to-end solution, including an encoder-decoder architecture and attention mechanism, and contains three steps: 1) from a sequence of ultrasound tongue image recordings, a 3D convolution neural network predicts the 93-dimensional embedding inputs of the pre-trained Tacotron2 model, 2) the Tacotron2 model converts this intermediate representation to a 80-dimensional mel-spectrogram, and 3) the WaveGlow model is applied for final inference. We demonstrated that the synthesized speech quality is significantly more natural with the proposed solutions than with our earlier model.

The code is accessible at \url{https://github.com/BME-SmartLab/UTI-to-STFT-Tacotron2}.

\section{Acknowledgements}

The research was partly supported by the European Union’s Horizon 2020 research and innovation programme under grant agreement No.\ 825619 (AI4EU), by the National Research Development and Innovation Office of Hungary (FK 124584 and PD 127915 grants; APH-ALARM / 2019-2.1.2-NEMZ-2020-00012 project) and through the Artificial Intelligence National Laboratory Programme. The Titan X GPU used was donated by NVIDIA Corporation. We would like to thank the subjects for participating in the listening test. 


\bibliographystyle{IEEEtran}

\bibliography{ref_collection_csapot_nourl}

\end{document}